\documentstyle[pre,aps,psfig]{revtex}

\begin{document}
\title{\bf A Cellular Automaton Model for Bi-Directional Traffic}
\author{ P. M. Simon $^{\star\dag}$ and H. A. Gutowitz $^{\star}$\\}
\address{
Ecole Sup\'erieure
de Physique et Chimie Industrielles, Laboratoire d'Electronique, 10 rue Vauquelin,
75005 Paris $^{\star}$,\\
TRANSIMS, TSA-DO/SA, MS-M997, Los Alamos National Laboratory, Los Alamos, NM,
87545, USA $^{\dag}$.
{\tt (simonp,hag)@neurones.espci.fr}}

\twocolumn
\draft
\maketitle

\begin{abstract}

We investigate a cellular automaton (CA) model 
of traffic on a bi-directional two-lane road.
Our model is an extension of the one-lane CA model of \cite{Nagel.92.Schreck},
modified to account for interactions mediated by passing, 
and for  a distribution of vehicle speeds. 
We chose values for the  various parameters to approximate
the behavior of real traffic. The  density-flow diagram for
the bi-directional model is 
compared to that of a one-lane model, showing the interaction of
the two lanes. Results were also compared
to experimental data, showing close agreement.
This model helps bridge the gap between
simplified cellular automata models
and the complexity of real-world traffic. 

\end {abstract}

\section{Introduction }

A variety of cellular automaton  models give a
discrete  approximation to traffic flow. Nagel and Schreckenberg's
\cite{Nagel.92.Schreck} one-lane model
accurately describes some of the main features of real traffic.
This lead to multi-lane \cite{rickert.96.etc.twolane}
and two-dimensional (e.g. \cite{Cuesta.etc,nagatani93})
CA models, which 
extended the range of traffic phenomena
which can be treated by cellular automata.
This paper continues in this line by introducing
a CA model in which vehicles move on two lanes, in opposite directions.
We find that even complicated interactions, such as occur during passing
into oncoming traffic, can be represented in a CA model.
This  supports the view that cellular automata provide a real
alternative to differential equations for the modeling of traffic
flow.

Multi-lane models provide rules for lane changes,
and we observe non-trivial interactions between the lanes.
With two lanes in  opposite directions,
these interactions depend strongly on the relationship
between the densities 
on the two lanes. When either or both of
the densities are large, interactions can be ignored since passing
is impossible; our model is equivalent to two copies of the standard
one-lane model.
If there are  no vehicles in the on-coming
lane, then our model behaves like  an asymmetric 
model with two lanes in the same direction. If there are only 
a few vehicles in a given lane and an intermediate number in the
oncoming lane, then vehicles in the oncoming lane will pass, and
slow progress on the given lane.

The one-lane CA model of  \cite{Nagel.92.Schreck} assumes that
all vehicles have the same maximum velocity. Indeed, even
a multi-velocity model quickly settles down to a state where
all vehicles move with the lowest speed;
the model unrealistically creates  platoons of vehicles, each
following a slow car.
As will be seen below, a distribution of speeds is especially
important for realism in the bi-directional model.

\section{Existing Models}

\paragraph{One Lane Model}
In the one-lane CA model of
\cite{Nagel.92.Schreck} each site may be
empty or occupied by a car with an integer velocity   
$v \in \{0 \ldots v_{max}\}$. 
$v_{max} = 5$ or greater gives good agreement with physical experiments.
The variable $gap$ gives the number of unoccupied sites in front of a vehicle.
$p_{decel}$ is the probability to randomly decelerate, and
$rand$ is a random number between 0 and 1.
One iteration consists of the three  following sequential steps, which are
applied in parallel to all cars:

\begin{enumerate}
\item
 Acceleration of free vehicles:
IF $(v < v_{max} )$ THEN $v=v+1$ 
\item
Slowing down due to other cars:
IF $(v > gap)$ THEN $v = gap$
\item
stochastic driver behavior:
IF $(v > 0)$ AND $ ($ rand $< p_{decel}$) THEN $v=v-1$ 
\end{enumerate}

These simple conditions already give realistic results.

\paragraph{Two-Lane Uni-directional Model}

The two-lane uni-directional model is built from two
parallel single-lane models \cite{rickert.96.etc.twolane}.
Four additional conditions govern the exchange of
vehicles between lanes.
First the vehicles change lanes, then the
one-lane algorithm is applied. 
This model introduces several new variables:
$gap_{same}(i),gap_{opp}(i)$: the number of unoccupied sites
in front of vehicle $i$ on the (same, opposite) lane, respectively.
$gap_{behind}(i)$ the number of unoccupied sites
{\it behind} the vehicle, on the opposite lane;
$l_{same}, l_{opp}$, $l_{back}$: the minimum free distance needed for a pass,
ahead on same lane, ahead  and behind on the opposite lane;
$ p_{change}$: the probability to change lanes.  The added rule is as following:

\begin{enumerate}
\item[4.]
IF ($ gap_{same}(i) < l_{same}$) AND
 $ (gap_{opp}(i) > l_{opp}$) AND
 $ (gap_{behind}(i) > l_{back}$) AND
 $ (rand < p_{change} $) THEN change lane
\end{enumerate}

This model has a symmetric and an 
asymmetric version. In the asymmetric version, there is
no passing on the right.

\section{Two-Lane Bi-directional Model}

The bi-directional model has several types. Either
passing is allowed on both lanes, only one or the other of the
lanes, or no passing is allowed.
Mixtures of these types are possible;  one type 
may be applied on a given section of roadway, and another
type on an adjacent section.

Consider a type in which passing is allowed. An algorithm
which controls passing must account for a number of
circumstances.
First, a vehicle must not decelerate
while passing; $p=0$ during the pass.
Moreover, if an on-coming car is seen, the passing car
must immediately return to its own lane. In reality
vehicles generally do not attempt to pass unless
the  pass can be completed; a model should reflect this.
We let  each vehicle measure the "local density": the density of
cars in front of the vehicle it would like to pass.
If the local density is sufficiently low, the vehicle has a good
chance of completing a pass, and we allow it to try.
Few passes will occur
when the global density is high,
even if the density is low on the on-coming lane. The lanes
become effectively de-coupled.

Three kinds of traffic jams can appear on a bi-directional road.
The most common is a start-stop
wave on one of the two lanes. Rarely, a jam is caused by
an audacious  driver tries  to pass though there is
no space to return to the home lane.
Finally, a jam may  occur when each of an adjacent pair of drivers,
one on each lane,
tries to pass simultaneously. Unless the symmetry between lanes is
broken (see below), this "super jam" halts all 
traffic on both lanes.

Our CA operates on a lattice with
two lanes in opposite directions (see figure 1). Each site has a state
$v \in \{-(v_{max}+1) \ldots  v_{max} +1 \} $, where 0 represents the
absence of a vehicle, $\pm 1$ a stopped vehicle, $\pm 2$ a vehicle moving
with speed 1 in the positive, resp. negative direction and so on.

Vehicle movement is calculated in a two-step process, following
\cite{rickert.96.etc.twolane}. First vehicles change lanes, 
then they advance.

In addition to the functions and variables introduced above,
we will need the following:
$v_{same}, v_{opp}$: the velocity of the vehicle
ahead on the (same, opposite) lane.
$H$:  true iff the vehicle is on its home lane;
$oncoming$: true iff $sign(v_{same}) \ne sign(v) $;
$l_{pass}$: if $gap_{same}  < l_{pass}$   AND $H$ then a
pass may be  attempted.
$l_{back}$: the distance a driver looks back for obstacles on the passing lane.
$l_{security}$: if $gap_{same}  < l_{security} $  AND $not(H)$ then 
the vehicle returns immediately to its home lane.
$D_L$: local density: the fraction of the $l_{density} = 2 \times v_{max} + 1$
sites  in front of the given
vehicle which are occupied;
$D_{limit}$: the maximum local density for a safe pass.

Space1: true if ($gap_{same}<l_{pass}$) AND ($gap_{opp}>l_{security}$)
AND ($gap_{behind}>l_{back}$).

Space2: true if ($gap_{opp}>l_{security}$) AND ($gap_{behind}>l_{back}$).

In our simulations we use the following values throughout:
$l_{pass} = v$, $l_{back} = v_{max}$, $l_{security} = 2 \times v_{max} +1$, $D_L$ = 
$\frac{2}{l_{density}}$,
$p_{change}$ = 0.7, and $p_{decel}$ =0.5.

\begin{figure}[t]
\centerline{\hbox{
\psfig{figure=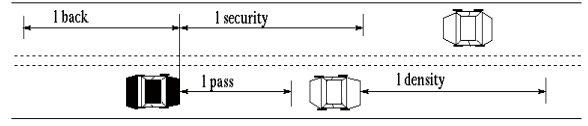,height=1.0in,width=3.2in}}}
\caption{bi-directional model}
\end{figure}

Lane changes are determined by the following steps:

\begin{enumerate}
\item
 IF  ($ H $ AND space1 AND 
  ($D_L \leq D_{limit}$)  AND
 ($rand < p_{change}$))   THEN change lane 
\item
IF ($not(H)$ AND ($(gap_{same}<l_{security})$ OR Space2))
 THEN change lane
\end{enumerate}

The first condition affects vehicles on their home lane. If a 
vehicle is in front of them, of the same sign, and  at a distance less than $l_{pass}$,
then they would like to pass that vehicle. However, a pass will
only be initiated if there is room far enough ahead on the
passing lane, and the number of cars in front of the vehicle
it would like to pass is small. Passing occurs
randomly, even if all these conditions are met,
the probability of changing lanes is
denoted $p_{change}$. The second condition concerns
vehicles in the midst of passing. They return to their home lane
if forced to by an oncoming vehicle, or if there is
space enough on the home lane that they can return without
braking.

Forward motion of a vehicle is determined as follows:

\begin{enumerate}
\item
IF  ($ |v|  \not=v_{max}$) THEN $v=v+sign(v)$
\item
IF  $ ((oncoming)$ AND ($ gap_{same} \leq (2\times v_{max} -1)$))
THEN $v = \lfloor gap_{same}/2 \rfloor$ 
\item 
IF ( $(not(oncoming)$) AND ($|v| > gap_{same}$)) THEN
$v = sign(v) \times gap_{same}$
\item
IF  (( $H$ AND ($|v|>1$) AND ($ rand < p_{decel} $) AND $not(oncoming)$)  THEN $v=v-sign(v)$
\item 
IF ( $H$ AND $(oncoming)$ AND ($|v|>1$) ) $v=v-sign(v)$
\end{enumerate}

These rules  (1) accelerate
the vehicle to maximum velocity, (2) rapidly decelerate  the vehicle if there
is an oncoming car too close, (3) decelerate the vehicle if it is
closing in on another, both in their home lane, and 
(4) randomly decelerates the vehicle
if it is on its home lane; if it is passing, it never decelerates
randomly. Finally, (5) breaks the symmetry between
the lanes, and thus prevents the emergence of a super jam.

\section{Results}

\begin{figure}[t]
\centerline{\hbox{
\psfig{figure=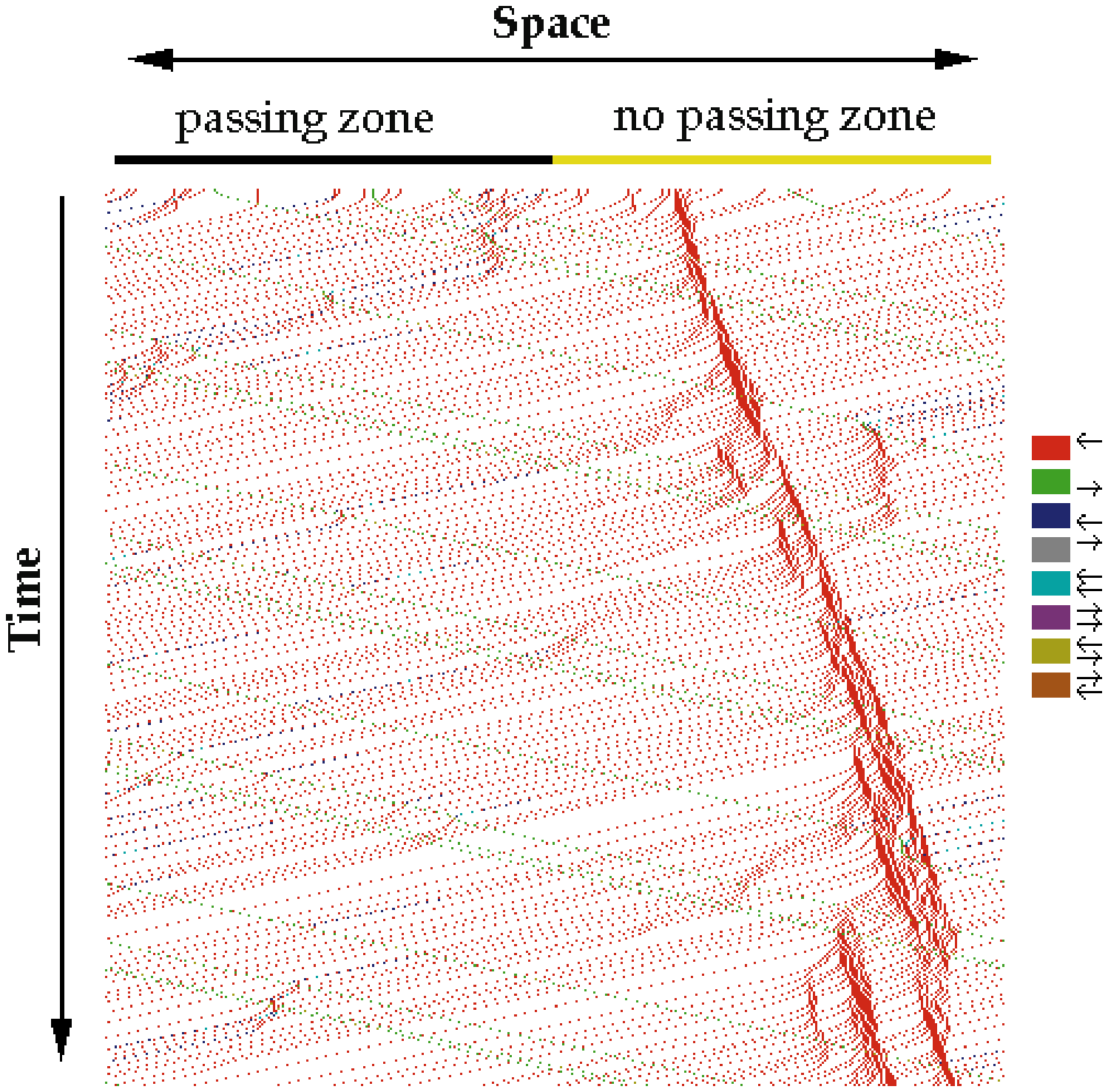,height=2.5in,width=3.5in}}}
\caption{passing and no-passing zones. \newline
A cell-wide section of roadway containing only one 
vehicle moving (rightward  resp. leftward) on its home lane is labeled 
red resp. green).
Other colors
give other possible situations, such as a single vehicle passing, going
rightward (violet). On the left-going lane the density is near critical 
for jams (0.1) and the density is low (0.01) on the right-going lane
(500 cells, periodic boundary conditions).
The route is divided into a passing zone (left half of the figure)
and a no-passing zone (right half of the figure).
Jams occur much more frequently in the no-passing zone.
All vehicles have the same maximum velocity of 5.}
\label{fig:nopass}
\end{figure}

The usual way to represent the behavior of a CA 
is with a space-time diagram, where space is represented on
the horizontal axis, and time on the vertical axis, with time proceeding
downward.  A typical space-time diagram for this model
is shown in figure \ref{fig:nopass}.

This figure
illustrates that passing can dramatically fluidity traffic.
The start-stop waves seen in the right half of the figure disappear
in the left half.
Here the densities have been chosen to maximize
the effects caused by interactions between  two lanes in opposite directions.

\begin{figure}[t]
\centerline{\hbox{
\psfig{figure=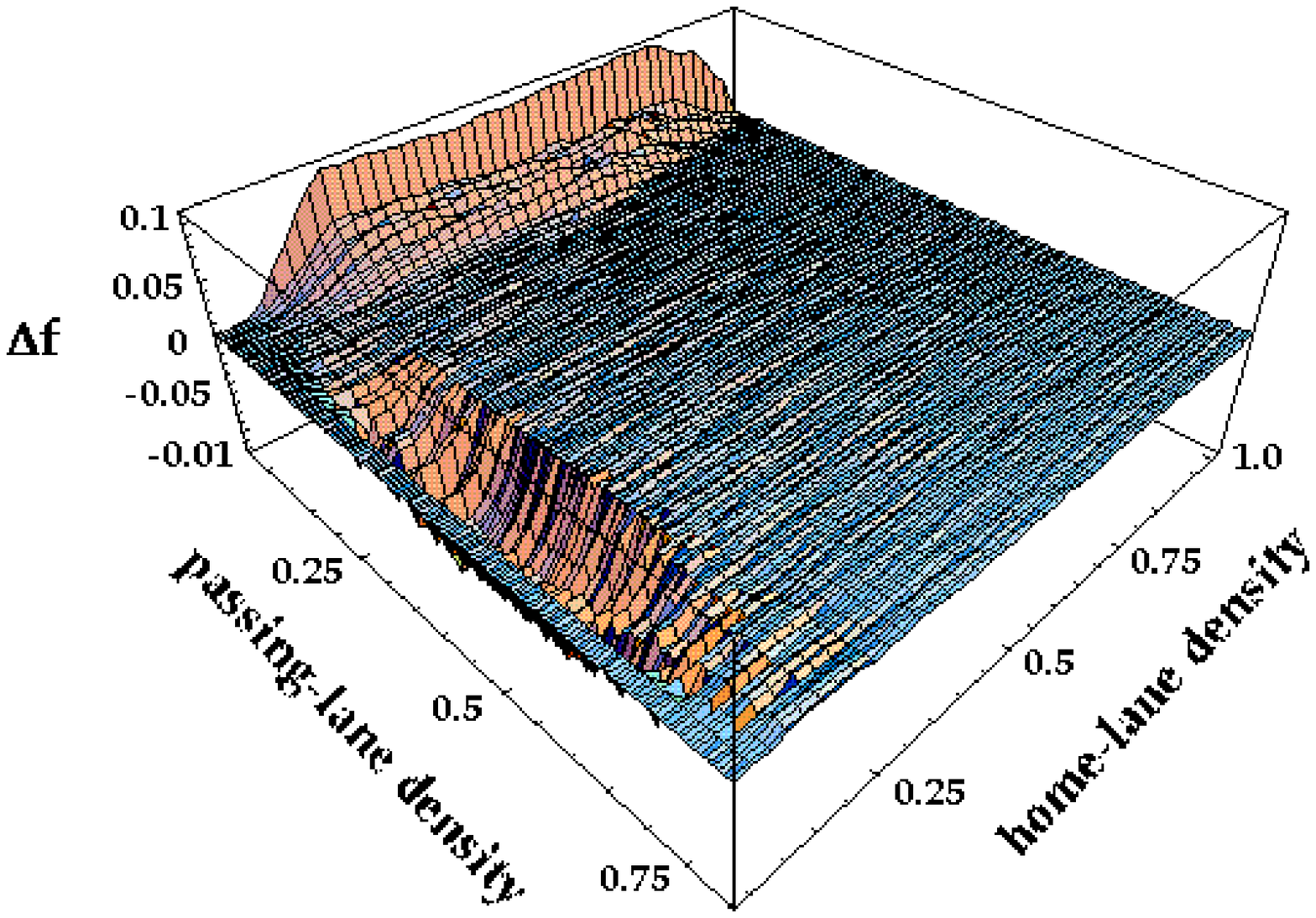,height=2.0in,width=2.7in}}}
\label{fig:contour}
\caption{3D plot of flow on the home lane.\newline
The height of the surface gives the difference between flow on 
the home lane in the bi-directional model and flow in a one-lane
model as a function of the densities on the home and passing lanes
of the bi-directional model.}
\end{figure}

We have explored the entire range of densities on the two lanes,
as shown in figure \ref{fig:contour}. Here the difference between
the flow on the home lane in 
two-lane model with passing on both lanes is compared to the flow
in a one-lane model ($\Delta f$).  When the density on either or both lanes
is large then there is little difference between the two-lane and
one-lane models. When the density on the passing lane
is small ( $  < 0.1 $), then the flow on the home
lane can be much greater than in a one-lane model.
blue). Maximum improvement occurs near 0 density on the passing 
lane. If the density on the home
lane is small ( $ < 0.25 $) then the flow may be {\it lower} than
in the corresponding one-lane model since when oncoming cars pass
other oncoming cars they can impede traffic on the home lane.
In an asymmetric
model in which only vehicles on the home lane can pass, this slowing
effect disappears.
This experiment was performed with a value of $p_{decel}$ of 0.5
and $p_{change}$ of 0.5.

\begin{figure}[t]
\centerline{\hbox{
\psfig{figure=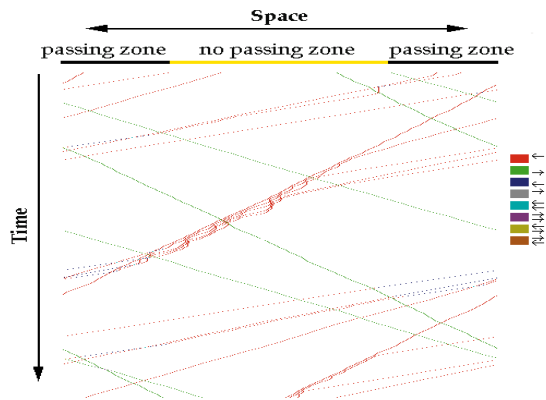,height=2.5in,width=3.5in}}}
\caption{A space-time diagram generated by  a multi-speed model}
Each car has a maximum velocity, uniformly distributed
between 2 and 5. In the center half of the figure no-passing
is allowed, while on the two sides passing is allowed.
Slower vehicles trace steeper lines. Note in particular
the large knot of cars near the center of the figure formed
by a slow (left-going) car entering a no-passing zone.
As the knot passes the leftmost boundary of the no-passing
zone, it disperses.
\label{fig:multi-speed}

\end{figure}

Our bi-directional model allows us to treat a host of
new phenomena. For instance, the multi-speed variant 
(figure \ref{fig:multi-speed}) can produce a complicated knot
of interactions resulting from   a slow car leading
a group of others through a no-passing zone. While in the
no-passing zone the faster cars bunch up behind the
slow car, forming a platoon.  When the platoon
reaches the end of the no-passing zone the platoon
disperses.

\section{Comparison with experiment} 

\begin{figure}[t]
\centerline{\hbox{
\psfig{figure=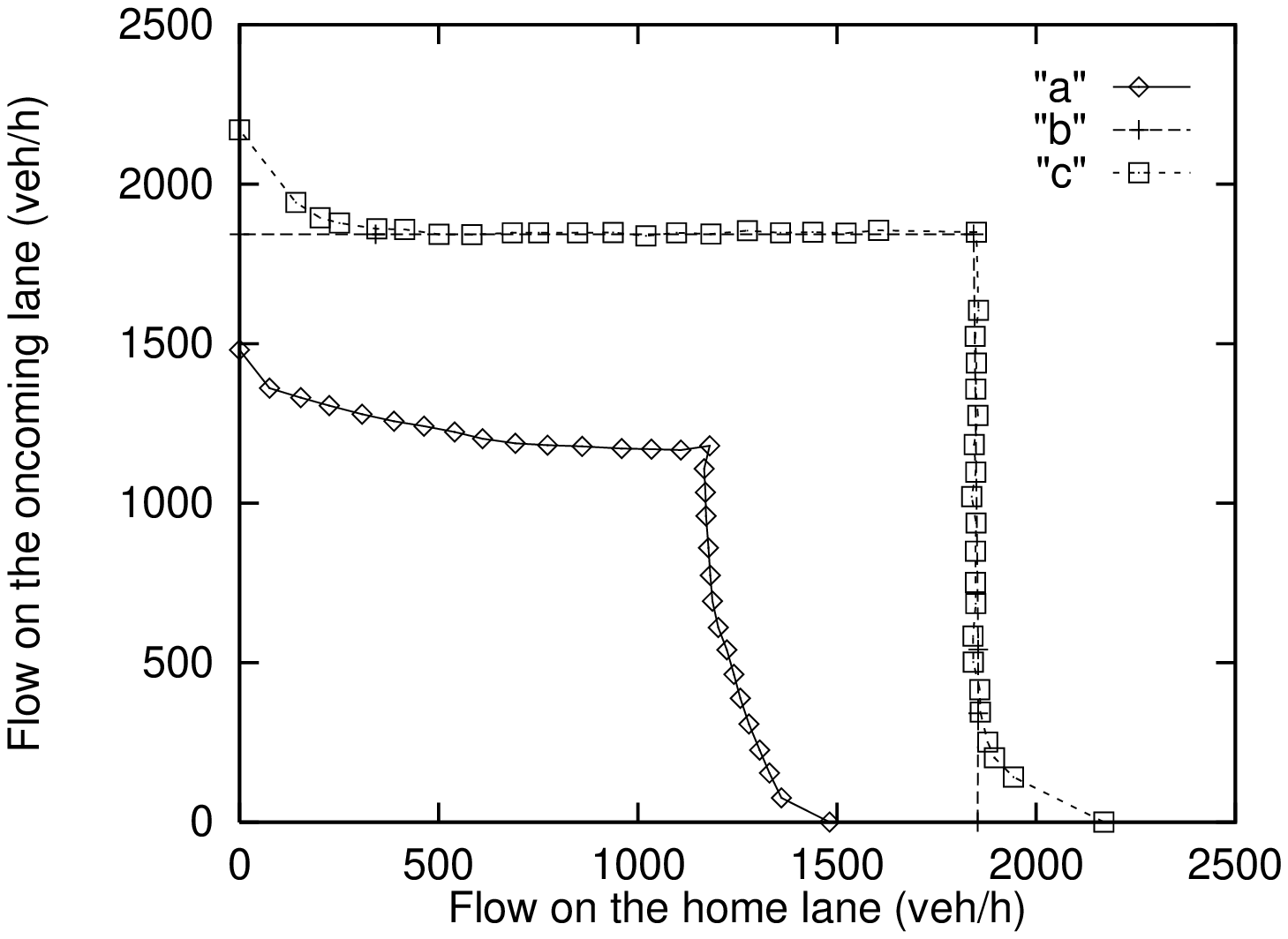,height=1.45in,width=1.8in}
\psfig{figure=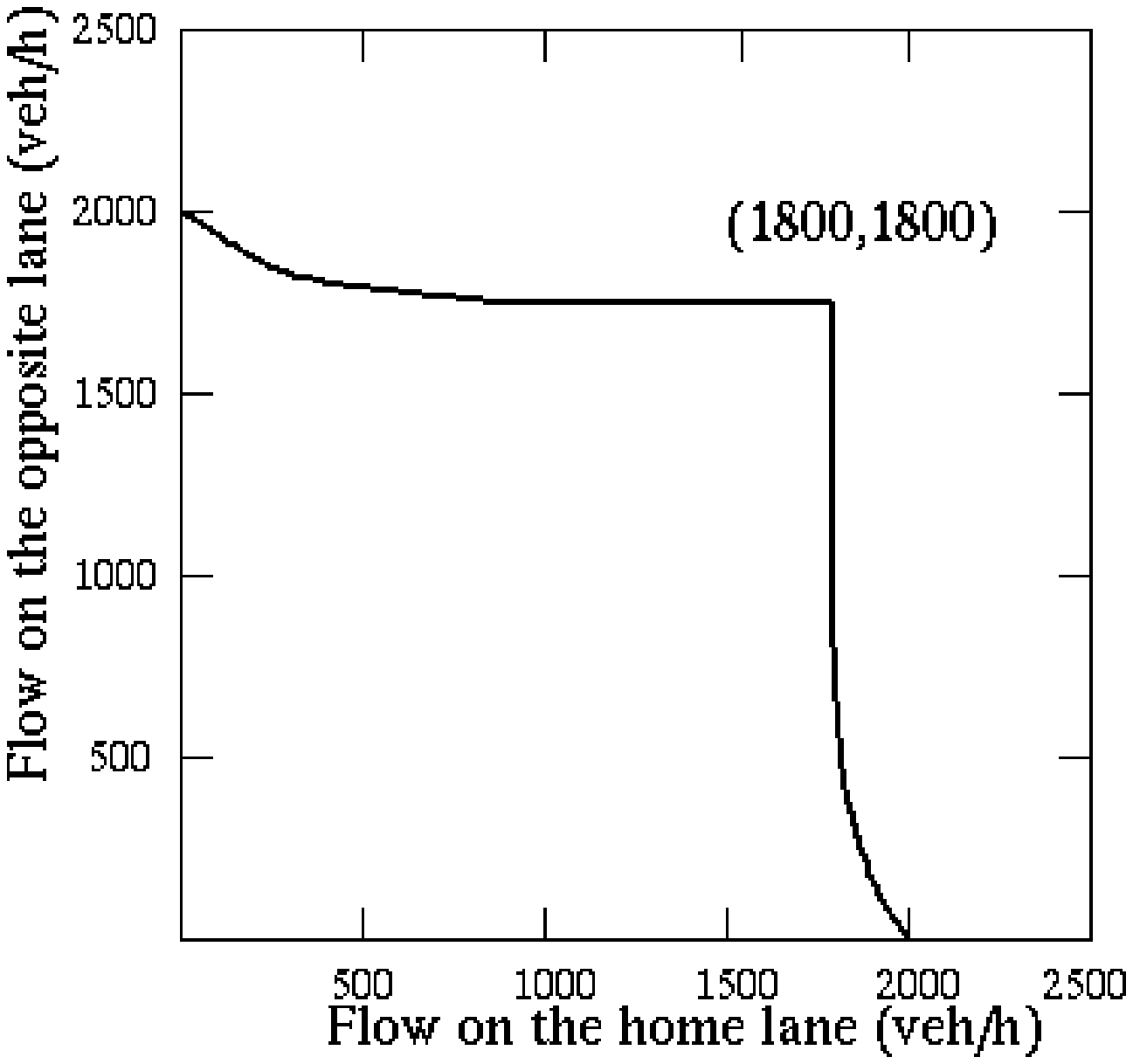,height=1.6in,width=1.8in}
}}
\caption{Comparison of simulation results with experimental data. \newline
left panel: The results of simulations under three conditions,
a)$p_{decel}=0.5 ,  p_{change}=0.5$,
b)$p_{decel}=0.25 ,  p_{change}=0$, and
c)$p_{decel}=0.25 , p_{change}=0.3$.
right panel: corresponding
physical measurements, modified from figure 5 of [5].} 
\label{fig:yag}
\end{figure}

In figure \ref{fig:yag}, we compare simulation results (left panel)
with the measurements of M. Van Aerde and S.Yagar
\cite{yagar} (right panel).
These experimental data  are supported by traffic
volume data from Australia and Canada
\cite{yagar}. Van Aerde and Yagar measured 
the maximum flow which can be obtained on the home lane,
as a function of the on-coming flow on the passing lane.
We made comparable measurements on our simulations, with
a variety of parameter setting, so as to approximate
the parameter values implicit in the experimental data.

In the simulations, we obtain maximum total flow when vehicles
are distributed uniformly, that is, half in each direction. The
simulation curves are obtained by varying the total number of
vehicles.

In both physical experiment and computer simulation, when densities are high,
flows are not affected by passing, resulting in a corner in the graph of maximum flow.
When one of the lane densities is fixed to a small enough value, the maximum flow 
increases on the other lane  due to passing.
Note that while our simulations are in close qualitative agreement 
with these experimental results of \cite{yagar}, these same results 
are {\it not} in agreement with the results reported in  
the previous \cite{HCM1} and current{\cite{HCM2}
editions of the influential Highway Capacity Manual (HCM).
The previous HCM\cite{HCM1} presents  a straight line of slope -1,
while the current HCM \cite{HCM2}
presents a  smoother curve, which suggests
that flows are correlated at high densities. Our simulation,
built from close modeling of microscopic interactions, gives theoretical
support for the view that the HCM's treatment of bi-directional
traffic should be revised in favor of the results of
Van Aerde and Yagar.

\section{Discussion}

We have shown that complicated interactions between
vehicles traveling on two-lane rural roads can be largely accounted
for by a few simple rules. By small extension of previous 
work, we have greatly increased
the realism which can be expected from cellular automaton models
of vehicular traffic. Traffic in the two-lane bi-directional
model approximates real traffic on both a microscopic and
macroscopic level. Indeed, our results suggest 
revision of a standard text in highway studies.

Though systematic exploration of the parameter space
should surely be undertaken, our model is robust
to small changes in parameter value.
Throughout the experiments described in this paper,
we have fixed the values of most parameters
at reasonable values, and focused on the
two most important variables: the density of vehicles on each lane.
We have localized the regions where lane interactions are strongest,
and explored how these interactions occur.
Due to the high relative speeds involved, these interactions have
practical importance in traffic safety.


\begin{thebibliography}{1}

\bibitem{Nagel.92.Schreck}
K. Nagel and M. Schreckenberg, J. Phys.~I France {\bf 2},  2221  (1992).

\bibitem{rickert.96.etc.twolane}
M. Rickert, K. Nagel, M. Schreckenberg, and A. Latour, Physica A {\bf 231},
  534  (1996).

\bibitem{Cuesta.etc}
J.~A. Cuesta, F.~C. Mart\'{\i}nez, J.~M. Molera, and A. S\'anchez, Phys. Rev. E
  {\bf 48(6)},  R4175  (1993).

\bibitem{nagatani93}
T. Nagatani, Phys. Rev. E {\bf 48},  3290  (1993).

\bibitem{yagar}
S. Yagar, Australian Road Research {\bf 13(1)},  3  (1983).

\bibitem{HCM1}
Highway Capacity Manual 1965. HRB Special report 87. HRB, National Research
  Coucil, Washington, D.C.  397 pp  (1965).

\bibitem{HCM2}
Highway Capacity Manual 1994, Special Report 209, Third Edition, TRB, National
  Research Coucil, Washington, D.C.  chapter 8 p6  (1994).

\end{thebibliography}
\end{document}